\documentclass[%
 reprint,
 superscriptaddress,
 amsmath,amssymb,
 aps,
 pre
]{revtex4-2}

\usepackage{physics}
\usepackage{graphicx}
\usepackage{bm}
\usepackage{amsmath}
\usepackage{hyperref}
\begin{document}

\title{Topological Flux on a Context Manifold Generates Nonreciprocal Collective Dynamics}

\author{Jyotiranjan Beuria}
\affiliation{IKS Research Centre, ISS Delhi, India}
\author{Venkatesh H. Chembrolu}
\affiliation{IKSMHA Centre, Indian Institute of Technology, Mandi, India}

\date{\today}

\begin{abstract}
Non-reciprocal interactions, where the influence of agent $i$ on $j$ differs from that of $j$ on $i$, are fundamental in active and living matter. Yet, most
models implement such asymmetry phenomenologically. Here we show that non-reciprocity can emerge from internal topology alone. Agents evolve 
on an internal ``context manifold'' coupled to a Chern-Simons gauge field. Because the gauge field is first order in time, it relaxes rapidly; eliminating 
it yields an effective transverse, antisymmetric interaction kernel that generically produces chiral waves, persistent vorticity, and irreversible 
state transitions. Numerical simulations reveal clear signatures of broken reciprocity: long-lived vortex cores, finite circulation, asymmetric information flow, and a nonzero reciprocity residual. The dynamics further exhibit pronounced hysteresis under parameter sweeps, demonstrating memory effects that cannot occur in reciprocal or potential-driven systems. These results identify Chern-Simons gauge fields as a minimal and universal source of directional influence and robust non-reciprocal collective behavior.
\end{abstract}

\maketitle

\section{Introduction}

Active matter systems consist of self-driven units that continuously consume energy and generate forces and flows at the mesoscale, giving rise to phenomena ranging from flocking \cite{sumpter2010collective,vicsek2012collective,popkin2016physics} and turbulence to pattern formation and topological modes \cite{marchetti2013hydrodynamics,bechinger2016active,shankar2017topological,shankar2022topological}. 
One of the characteristic features of active systems is the emergence of nonreciprocal interactions \cite{knevzevic2022collective, dinelli2023non, fruchart2021non}, which break time-reversal symmetry and enable sustained currents, circulating flows, and chiral collective motion \cite{han2021fluctuating,shankar2022topological}. 
Recent work has shown that such non-equilibrium interactions can be captured through antisymmetric couplings, including odd elasticity \cite{scheibner2020odd}, Lorentz-like forces, and Hall stresses \cite{lou2022odd}, that produce robust rotational dynamics and topologically protected behaviors.

Despite this progress, most active-matter models treat nonreciprocal interactions as acting directly in physical space or through simple internal degrees \cite{ziepke2022multi} of freedom such as orientation or spin. In many systems, however, particles also carry auxiliary variables—such as polarity \cite{baconnier2025self}, phase \cite{o2017oscillators}, or chirality \cite{liebchen2017collective}, that evolve on low-dimensional manifolds and modulate their interactions. Rather than constituting mere auxiliary labels, such internal coordinates can carry intrinsic geometric and topological structure, shaping how information, influence, or bias is transmitted across the population. We therefore hypothesize that active agents may inhabit an internal state space with nontrivial geometry whose dynamics feed back onto motion in physical space. How generic features of this internal geometry, independent of microscopic details, can give rise to emergent collective behavior remains largely unexplored. Addressing this question is particularly relevant for living active matter, such as bird flocks, where coordination appears to be mediated by internal state variables beyond purely spatial degrees of freedom.

We introduce a minimal gauge-theoretic framework in which the internal states of active agents evolve on an internal manifold endowed with a Chern-Simons (CS) gauge structure, while their positions evolve in physical space. 
The CS action produces antisymmetric, nonreciprocal interactions in the internal manifold, the underlying flux generates transverse drift forces on the agents in physical space. 
This mechanism yields circulating currents, rotating clusters, and large-scale chiral patterns, even in the absence of explicit alignment or metric interactions in real space.



In our formulation, the internal manifold is referred to as a “context space” to emphasize that it encodes the instantaneous internal configuration in which interactions are evaluated. The term “context” is used in a purely structural sense: it labels an internal coordinate system, distinct from physical position, whose geometry shapes how nonreciprocal interactions are mediated. Thus, each point in context space represents an equivalence class of internal conditions under which otherwise identical interaction rules can yield different effective forces. In this view, context space functions as an intrinsic coordinate chart that organizes the internal state of the system and determines how interaction asymmetries are expressed.

A central modeling assumption in our framework is a natural separation of time
scales between the slow evolution of the context density and the rapid
relaxation of the Chern--Simons gauge field. The gauge sector is first order in
time and contains no Maxwell-type kinetic term; it therefore carries no inertia
and adjusts essentially instantaneously to changes in the density. By contrast,
the density reorganizes on a much slower collective time scale, allowing the
gauge field to be slaved to the instantaneous density configuration and
reducing the dynamics to an overdamped force--balance relation. Similar
slow--fast reductions appear in Hall and magnetized fluids, where
electromagnetic fields equilibrate quickly relative to the bulk flow
\cite{avron1995viscosity}, and in active or chemically
mediated systems, where eliminating fast mediator fields yields effective
non-reciprocal interactions \cite{saha2020scalar,fruchart2021non}. 

Eliminating
a fast, first-order field generically produces reduced equations that mix
antisymmetric (gyroscopic) and symmetric components, even when the underlying
Lagrangian is fully conservative. Consequently, the effective dynamics obtained
after gauge-field elimination need not remain reciprocal, providing a natural
route to the persistent circulation and chiral flow patterns observed in our
model. This mechanism is broadly consistent with the slaving principles of
synergetics \cite{haken2012advanced} and with classical derivations of
hydrodynamic reductions in pattern-forming systems \cite{cross1993pattern}.

The organization of this work is as follows. In section \ref{sec:context}, we define what context means through equivalence relations. In section \ref{sec:cs}, we introduce the Chern-Simons field dynamics. In section \ref{sec:cs_conti}, we present a continuum context dynamics and its Fourier transformation which plays an important role for faster computation. In section \ref{sec:physical}, we present the procedure for mapping context space dynamics to physical space dynamics. Section \ref{sec:numerical} presents numerical simulations and associated results. We conclude in section \ref{sec:conclusion}.

\section{Context manifold, equivalence classes, and context modes}
\label{sec:context}
Active agents often possess internal variables that modulate how they respond to neighbors. 
When interactions depend only on relative internal conditions rather than absolute values, 
many distinct internal configurations become indistinguishable from the point of view of the interaction rules. 
This induces a natural equivalence structure on the internal state space.

Let $x \in \mathcal{X}$ denote a raw internal state. 
Two internal states $x$ and $x'$ are defined to be interaction-equivalent if they generate identical responses to all neighboring agents,
\begin{equation}
    x \sim x'
    \quad \Longleftrightarrow \quad 
    \mathcal{K} (x,y)=\mathcal{K} (x',y) \;\;\text{for all neighbors } y ,
\end{equation}
where $\mathcal{K}(x,y)$ is the interaction kernel governing the influence of a neighbor in state $y$ on an agent in state $x$.  
This relation partitions $\mathcal{X}$ into context classes $[x]$, each containing all internal states that are operationally indistinguishable under the interaction rules.

The space of all context classes defines a reduced internal coordinate space,
\begin{equation}
    \mathcal{C} = \mathcal{X}/\!\sim,
\end{equation}
which we refer to as the context manifold.  
A coordinate $c \in \mathcal{C}$ parametrizes these classes and will be termed a context mode.  
The context manifold therefore captures only those internal distinctions that have dynamical consequences; all redundant internal variability has been quotiented out.

A simple and instructive example arises when each agent carries two periodic internal variables
\begin{equation}
    x = (\theta_1,\theta_2) \in [0,2\pi)\times[0,2\pi),
\end{equation}
which might represent oscillatory phases or cyclic internal modes.  
If interactions depend only on phase differences,
\begin{equation}
    \mathcal{K}(x_i,x_j) 
    = f(\theta_{1,i}-\theta_{1,j},\;\theta_{2,i}-\theta_{2,j}),
\end{equation}
then shifts by full $2\pi$ cycles leave all interaction responses unchanged.  
States differing by integer multiples of $2\pi$ in either coordinate therefore fall into the same context class,
\begin{equation}
    (\theta_1,\theta_2)
    \sim (\theta_1+2\pi n_1,\;\theta_2+2\pi n_2),
    \qquad n_1,n_2 \in \mathbb{Z}.
\end{equation}
The quotient of the square $[0,2\pi)\times[0,2\pi)$ under these identifications is the two-torus
\begin{equation}
    \mathcal{C} \cong S^1 \times S^1 = T^2,
\end{equation}
with the pair $c=(\theta_1,\theta_2)$ serving as the context mode. Thus,
This construction illustrates how the geometry and topology of the context manifold arise directly from the symmetries and invariances of the interaction kernel, rather than from any ad hoc modeling choice.

In the following section we formulate the dynamics on $\mathcal{C}$ using a gauge-theoretic description. 
The antisymmetric, nonreciprocal component of interactions emerges from the geometry of the context manifold through a Chern-Simons connection whose curvature encodes how context modes vary across $\mathcal{C}$.

\section{Chern-Simons dynamics on context space}
\label{sec:cs}
We now consider a population of active agents whose dynamics involve both physical positions and an auxiliary internal coordinate $c\in\mathcal{C}$, where $\mathcal{C}$ is a smooth internal manifold. This internal coordinate represents the local context in which interactions are evaluated, for example, an internal orientation, phase, or latent mode that modulates each agent’s response. We refer to $\mathcal{C}$ as a ``context manifold" to emphasize that it supplies the internal conditions under which effective interactions are defined, without introducing any additional physical degrees of freedom. Each agent carries a scalar field $\rho(c,t)$ defined on $\mathcal{C}$, representing a density or internal drive associated with that point in the manifold. To encode nonreciprocal interactions within this internal space, we introduce a $U(1)$ Chern-Simons gauge field $a_\mu(c,t)$ on $\mathcal{C}\times\mathbb{R}$. 
\subsection*{The Chern-Simons action}
Let $a_\mu=(a_0,a_i)$, where $a_0$ and $a_i$ are temporal and spatial components of the Chern-Simons gauge field. The CS action \cite{dunne2002aspects} is
\begin{equation}
S_{\rm CS}
= \frac{\kappa}{4\pi}
\int \epsilon^{\mu\nu\rho} a_\mu \partial_\nu a_\rho \, d^2 c\, dt,
\label{CSaction}
\end{equation}
where $\epsilon^{\mu\nu\rho}$ is the usual anti-symmetric Levi-Civita tensor. Agents contribute a conserved current $j^\mu=(\rho,\,j^i)$, giving the coupling
\begin{equation}
S_{\rm coup} = -\int a_\mu j^\mu\, d^2c\,dt .
\end{equation}
The total action is the combination of both $S_{\rm CS}$ and  $S_{\rm coup}$, i.e., 
\begin{equation}
S[a_\mu]=S_{\rm CS} + S_{\rm coup}.
\end{equation}
The variation of total action $S[a_\mu]$ yields the CS equation
\begin{equation}
\frac{\kappa}{2\pi} \epsilon^{\mu\nu\rho} \partial_\nu a_\rho
= j^\mu .
\label{CSequation}
\end{equation}

The $\mu=0$ component gives the Chern-Simons Gauss law constraint
\begin{equation}
\frac{\kappa}{2\pi} B(c,t)
=
\rho(c,t),
\qquad
B=\epsilon^{ij}\partial_i a_j.
\label{eq:Gauss}
\end{equation}

$B(c,t)$ plays the role of an emergent ``magnetic'' flux on $\mathcal{C}$, measuring the local curl or 
circulation of the gauge field $a_i$.  This flux is not a physical magnetic field; rather, it 
encodes the geometric twisting or directional bias that agents experience when interacting through 
the internal coordinate $c$.  The Gauss law constraint then states that internal activation acts as a source of topological flux: wherever agents accumulate in context space, the interaction geometry acquires a corresponding amount of rotational bias. This induced flux is the origin of non-reciprocity in the effective interactions.

\subsection*{The Green's function and gauge field solution}

To solve the Gauss law it is useful to note that it does not determine the
full gauge field, but only its spatial curl.  Any gauge potential may be
decomposed as
\begin{equation}
a(c) = a^{(0)}(c) + a^{\rm hom}(c),
\label{eq:split}
\end{equation}
where $a^{(0)}$ is a particular solution of the Gauss constraint,
\begin{equation}
\frac{\kappa}{2\pi}\,\epsilon^{ij}\partial_i a^{(0)}_j = \rho,
\end{equation}
and $a^{\rm hom}$ is an arbitrary curl--free (homogeneous) field satisfying
\begin{equation}
\epsilon^{ij}\partial_i a^{\rm hom}_j = 0.
\end{equation}
Thus the Green's-function inversion of the Gauss law yields only the
particular piece $a^{(0)}$: it enforces the constraint but does not capture
the homogeneous component $a^{\rm hom}$, which is instead generated
dynamically by the second Chern--Simons equation (see Apendix section \ref{sec:Ampere-like-law}),
\begin{equation}
-\frac{\kappa}{2\pi}\,\epsilon^{ij}\partial_t a_j = j^i .
\label{eq:ji}
\end{equation}
The Gauss-law solution $a^{(0)}$ should therefore be understood as the
leading (zeroth-order) contribution to the full gauge field, with additional
corrections arising from the time-evolution of $a^{\rm hom}$.

Solving Eq.~\eqref{eq:Gauss} requires the Green function $G$ of the Laplacian.
If the context manifold $\mathcal{C}$ is compact (for example, a torus or any closed two–dimensional manifold), then the Laplacian $\Delta$ (also known as Laplace-Beltrami operator) possesses a constant zero mode and cannot be inverted directly. 
Accordingly, the Green's function must be defined with the zero-mode subtraction term $\frac{1}{|\mathcal{C}|}$. Here $|\mathcal{C}| = \int_{\mathcal{C}} d^{2}c$ denotes the total volume of the compact manifold. For example, 
on the flat two-torus $\mathbb{T}^2$, equipped with the standard Euclidean metric 
$g_{ij}=\delta_{ij}$, the Laplace--Beltrami operator reduces to the ordinary
two--dimensional Laplacian,
\[
\Delta f(c)
=
\partial_{1}^{2} f(c) + \partial_{2}^{2} f(c),
\qquad
c=(c_1,c_2)\in\mathbb{T}^2.
\]

We write the gauge field in 2D as the rotated gradient of a scalar
potential, $a_j = \epsilon_{ji}\partial_i \phi$, which implies
\begin{equation}
\epsilon^{ij}\partial_i a_j = -\Delta\phi.
\end{equation}
Hence Gauss law reduces to a Poisson equation on the torus,
\begin{equation}
-\Delta\phi(c) = \frac{2\pi}{\kappa}\,\rho(c).
\end{equation}
Because $\mathbb{T}^2$ is compact, the Laplacian has a constant zero mode,
and its Green's function is defined by
\begin{equation}
-\Delta G(c,c') = \delta(c-c') - \frac{1}{|\mathcal{C}|}.
\end{equation}
Inverting the Poisson equation gives
\begin{equation}
\phi(c)
=
\frac{2\pi}{\kappa}
\int_{\mathcal{C}} G(c,c')\,\rho(c')\,dc',
\end{equation}
and substituting $a_j=\epsilon_{ji}\partial_i\phi$ yields the particular
solution of the Gauss constraint,
\begin{equation}
a_j(c)
=
\frac{2\pi}{\kappa}
\int_{\mathcal{C}}
\epsilon_{ji}\,\partial_i G(c-c')\,\rho(c')\,dc'.
\label{eq:a0Solution}
\end{equation}

\subsection*{Origin of drift velocity and overdamped regime}
The drift $v$ in context space follows directly from the minimal
Chern--Simons coupling and is given by $v=-\gamma a$. Varying the coupling term $S_{\rm coup}$ with respect to the spatial
current gives the generalized force conjugate to motion,
\begin{equation}
    F_i \;=\; -\,\frac{\delta S_{\rm coup}}{\delta j^i}
    \;=\; -\,a_i .
\end{equation}
In the overdamped regime relevant for active systems, friction dominates
inertial effects and the velocity is proportional to the applied force,
\begin{equation}
    v_i = \gamma\, F_i .
\end{equation}
Here $\gamma$ controls the force in the context manifold. Subsequent numerical simulation will present a very crucial dynamic role of $\gamma$ related to hysteresis. Combining these relations yields a direct proportionality between drift and
the context--space gauge field,
\begin{equation}
    v_i = -\,\gamma\, a_i .
\end{equation}
Thus the Chern--Simons gauge potential acts as a Lorentz-like transverse
force field that sets the drift of each agent and establishes the dynamical
link between gauge structure and physical motion.
\subsection*{Emergence of reciprocity}
From particular solution of gauge field mentioned in equation \eqref{eq:a0Solution}, the agents experience a drift
\begin{equation}
    v^{(0)}_i(c)
    = 
    -\gamma\,a_i^{(0)}(c)
    =
    \int \mathcal{K}_i(c,c')\,\rho(c')\,dc'
    = \mathcal{K}_i * \rho,
\end{equation}
with interaction kernel
\begin{equation}
    \mathcal{K}_i(c,c')
    =
    -\frac{2\pi\gamma}{\kappa}\,
    \epsilon_{ij}\, \partial_j G(c-c').
    \label{eq:Kernel}
\end{equation}
and $\mathcal{K}_i * \rho$ denotes
\begin{equation}
    (\mathcal{K}_i * \rho)(c)
    = 
    \int_{\mathcal{C}} \mathcal{K}_i(c,c')\, \rho(c')\, dc' .
\end{equation}

Let us consider a simple of example of two agents located at context positions $c_1$ and $c_2$, the density is
\[
\rho(c') = \delta(c' - c_1) + \delta(c' - c_2),
\]
so the drift at $c_1$ and $c_2$ becomes
\[
v^{(0)}(c_1) = \mathcal{K}(c_1,c_2), \qquad
v^{(0)}(c_2) = \mathcal{K}(c_2,c_1).
\]
We therefore identify the pairwise contributions
\begin{equation}
v^{(0)}_{1\leftarrow 2} = \mathcal{K}(c_1,c_2), \qquad
v^{(0)}_{2\leftarrow 1} = \mathcal{K}(c_2,c_1).
\end{equation}
Because the kernel is antisymmetric,
\[
\mathcal{K}(c_2,c_1) = -\,\mathcal{K}(c_1,c_2),
\]
the zeroth-order interactions are strictly reciprocal,
\begin{equation}
v^{(0)}_{1\leftarrow 2} + v^{(0)}_{2\leftarrow 1} = 0,
\end{equation}
and thus,
\begin{equation}
F_{1\leftarrow 2} + F_{2\leftarrow 1} = 0,
\end{equation}

\subsection*{Emergence of non-reciprocity}
The zeroth-order dynamics generated by the Chern--Simons Gauss constraint
produce a velocity field $v^{(0)} = \mathcal{K} * \rho$ that is strictly
reciprocal.  Non-reciprocal interactions arise only when the full
Chern--Simons dynamics are taken into account.  Beyond the Gauss law, the
gauge field also couples to the spatial current (see equation \eqref{eq:ji}) $j^i = \rho\, v^i$ through
\begin{equation}
    -\,\frac{\kappa}{2\pi}\,\epsilon^{ij}\partial_t a_j
    \;=\; j^i
    \;=\; \rho\, v^i ,
    \label{CScurrent}
\end{equation}
so that, with $v^i = -\gamma\, a^i$, the gauge potential acquires an
autonomous, self-driven evolution,
\begin{equation}
    \partial_t a
    \;=\;
    \frac{2\pi\gamma}{\kappa}\,
    \epsilon \!\cdot\! (\rho\, a).
\end{equation}

Eliminating $a$ self-consistently leads to nonlinear density dependence
because the Chern--Simons evolution couples $\partial_t a$ to $\rho\,a$,
while the Gauss-law solution already gives $a^{(0)}\!\propto\!\rho$.
Iterating this feedback generates corrections $a^{(1)}\!\sim\!\rho^2$,
$a^{(2)}\!\sim\!\rho^3$, etc., so that $v=-\gamma a$ becomes a nonlinear
functional of~$\rho$.

The velocity is a nonlinear functional of the density obtained by
eliminating the gauge field self-consistently.  The leading terms read
\[
v(c)=v^{(0)}(c)+v^{(1)}(c)+\cdots,
\]
with
\[
v^{(0)}(c)=\int_{\mathcal C}\mathcal{K}(c,c_1)\,\rho(c_1)\,dc_1,
\]
and the first nonlinear correction given 
by the nested convolution
\[
v^{(1)}(c) = \int_{\mathcal C}\mathcal{K}(c,c_1)\,\rho(c_1)
\left[\int_{\mathcal C}\mathcal{K}(c_1,c_2)\,\rho(c_2)\,dc_2\right]dc_1.
\]
Equivalently,
\[
v=\mathcal{K}*\rho+\mathcal{K}*(\rho\;(\mathcal{K}*\rho))+\cdots,
\]
which makes explicit that the higher terms are nonlocal and involve
pointwise products of \(\rho\) with lower-order convolutions.  The
formal series is valid in the perturbative regime in which successive
terms are parametrically small.

Now going back to our example of two agents, because the corresponding
velocity correction $v^{(1)} = -\gamma\, a^{(1)}$ is not antisymmetric under
exchange of particle labels, the first-order pairwise contributions do not
cancel:
\begin{equation}
    v^{(1)}_{1} + v^{(1)}_{2} \neq 0
    \qquad \Longrightarrow \qquad
    F_{1\leftarrow 2} + F_{2\leftarrow 1} \neq 0 .
\end{equation}
Thus, it explains the origin of non-reciprocal interactions in active matter systems.

\section{Continuum Context Dynamics}
\label{sec:cs_conti}
To investigate the emergent structures predicted by the analytic 
framework, we simulate the continuum evolution of the context density 
$\rho(c,t)$ on the two--torus $\mathcal C = T^2$ and map the resulting 
velocity fields into physical space through the embedding 
$\Phi:\mathcal C\to\mathbb R^2$.  
The continuum formulation is the natural numerical setting for the 
Chern--Simons (CS) gauge theory: in the overdamped regime the gauge 
field carries no independent degrees of freedom, and the CS Gauss law 
reduces the velocity field to a nonlocal functional of the density.  
All gauge-field effects, including antisymmetric drift and higher-order 
nonreciprocal corrections, therefore enter through convolution 
operators that are evaluated spectrally.

\subsection{Continuum PDE formulation}

We now pass to a continuum description of the dynamics. While the preceding discussion focused on the kinematics of individual trajectories, a complete statistical formulation requires an explicit evolution equation for the density field. In the continuum limit, the density $\rho(c,t)$ evolves according to a conservation law supplemented by diffusive regularization,
\begin{equation}
    \partial_t \rho
    =
    -\nabla\!\cdot\!\bigl(\rho\,v\bigr)
    + D_c \Delta \rho,  
    \label{eq:PDE}
\end{equation}
where $v(c,t)$ denotes the effective velocity field generated by the interaction kernel, and $D_c$ is the context diffusion coefficient. The first term enforces local conservation of probability through advective transport, while the second term accounts for stochastic fluctuations and coarse-graining at finite resolution. This formulation provides the natural starting point for a Fourier-space analysis, which we employ below to elucidate the collective modes and nonreciprocal structure of the dynamics. We also know that
\begin{align}
    v
    &=v^{(0)}+v^{(1)}+\cdots\nonumber\\ 
    &=
    \mathcal K * \rho
    +
    \mathcal K * \!\bigl(\rho(\mathcal K*\rho)\bigr)
    +
    \cdots,
    \label{eq:VelocityExpansion}
\end{align}
where $\mathcal K$ is the antisymmetric kernel derived previously.
The usual Fourier conventions are as follow.
\[
\widehat{f}(k)=\int f(c)\,e^{-i k\!\cdot\! c}\,dc,
\qquad
f(c)=\frac{1}{(2\pi)^2}\int \widehat{f}(k)\,e^{i k\!\cdot\! c}\,dk,
\]
\[
\widehat{fg}(k)=\frac{1}{(2\pi)^2}\int \widehat{f}(q)\,\widehat{g}(k-q)\,dq,
\qquad
\widehat{f*g}(k)=\widehat{f}(k)\,\widehat{g}(k).
\]
Let \(C:=\dfrac{2\pi\gamma}{\kappa}\) and \(\varepsilon_{12}=1,\ \varepsilon_{21}=-1\).  
Fourier-transform of equation \eqref{eq:PDE} gives
\begin{equation}
    \partial_t\widehat{\rho}(k,t) = - i k_j\,\widehat{\rho v_j}(k,t) - D_c|k|^2\widehat{\rho}(k,t).
    \label{eq:delrho_kt}
\end{equation}
Also, Fourier transformation of equation \eqref{eq:VelocityExpansion} gives

\begin{align}
  \widehat{v^{(0)}_j}(p) &= -C\,i\,\frac{\varepsilon_{j a}\,p_a}{|p|^{2}}\;\widehat{\rho}(p), \\
  \widehat{v^{(1)}_j}(p) &= -\frac{C^{2}}{(2\pi)^{2}}\;
\frac{\varepsilon_{j a}\,p_a}{|p|^{2}}\;
\int
\frac{\varepsilon_{b c}\,(p-r)_c}{|p-r|^{2}}\;
\widehat{\rho}(r)\,\widehat{\rho}(p-r)\;d^{2}r.
\end{align}
We can express \(\widehat{\rho v_j}=\widehat{\rho v^{(0)}_j}+\widehat{\rho v^{(1)}_j}+\cdots\). Next, we calculate each of these contributions up to the cubic order in $\hat{\rho}(k)$.
The quadratic term comes from \(v^{(0)}\) as follows:
\begin{align}
\widehat{\rho v^{(0)}_j}(k)
&=\frac{1}{(2\pi)^2}\int \widehat{\rho}(q)\,\widehat{v^{(0)}_j}(k-q)\,d^2q \nonumber \\
&=\frac{1}{(2\pi)^2}\int \widehat{\rho}(q)\Big(-C\,i\frac{\varepsilon_{j a}(k-q)_a}{|k-q|^2}\widehat{\rho}(k-q)\Big)d^2q.
\end{align}
Multiplying by \(-i k_j\) gives
\begin{equation}
\left.\partial_t\widehat{\rho}(k)\right|_{\text{quad}}
= -\frac{C}{(2\pi)^2}\;k_j\int\frac{\varepsilon_{j a}\,(k-q)_a}{|k-q|^{2}}
\;\widehat{\rho}(q)\,\widehat{\rho}(k-q)\;d^2q.
\end{equation}
Next, the cubic term comes from \(v^{(1)}\) as follows:
\begin{equation}
\widehat{\rho v^{(1)}_j}(k)
=\frac{1}{(2\pi)^2}\int \widehat{\rho}(q)\,\widehat{v^{(1)}_j}(k-q)\,d^2q, 
\end{equation}
so substituting \(\widehat{v^{(1)}_j}\) and multiplying by \(-i k_j\) yields
\begin{align}
\left.\partial_t\widehat{\rho}(k)\right|_{\text{cubic}}
= i\,\frac{C^{2}}{(2\pi)^4}\;k_j
\int\int
\frac{\varepsilon_{j a}\,(k-q)_a}{|k-q|^{2}}\;\nonumber \\
\frac{\varepsilon_{b c}\,(k-q-r)_c}{|k-q-r|^{2}}\;
\widehat{\rho}(q)\,\widehat{\rho}(r)\,\widehat{\rho}(k-q-r)\;d^2r\,d^2q.
\end{align}

\begin{align}
\left.\partial_t\widehat{\rho}(k)\right|_{\text{cubic}}
&=
i\,\frac{C^{2}}{(2\pi)^4}\; k_j
\int\!\!\int
\frac{\varepsilon_{j a}\,(k-q)_a}{|k-q|^{2}}\;
\frac{\varepsilon_{b c}\,(k-q-r)_c}{|k-q-r|^{2}}
\nonumber\\[4pt]
&\qquad\times\;
\widehat{\rho}(q)\,
\widehat{\rho}(r)\,
\widehat{\rho}(k-q-r)\;
d^2 r\, d^2 q.
\end{align}

Combining quadratic and quadratic contributions, equation \eqref{eq:delrho_kt} for \(k\neq0\) becomes
\begin{align}
\partial_t\widehat{\rho}(k,t)
&=
-\frac{C}{(2\pi)^2}\, k_j
\int
\frac{\varepsilon_{j a}\,(k-q)_a}{|k-q|^{2}}\;
\widehat{\rho}(q,t)\,\widehat{\rho}(k-q,t)\,d^2 q
\nonumber \\[6pt]
&\quad
+\, i\,\frac{C^{2}}{(2\pi)^4}\, k_j
\int\!\!\int
\frac{\varepsilon_{j a}\,(k-q)_a}{|k-q|^{2}}\;
\frac{\varepsilon_{b c}\,(k-q-r)_c}{|k-q-r|^{2}}
\nonumber
\\[-2pt]
&\qquad\qquad\qquad
\times\;
\widehat{\rho}(q,t)\,
\widehat{\rho}(r,t)\,
\widehat{\rho}(k-q-r,t)\;
d^{2}r\, d^{2}q
\nonumber \\[6pt]
&\quad
-\,D_c\,|k|^{2}\,\widehat{\rho}(k,t).
\end{align}

\section{Mapping Context Dynamics to Physical Embedding}
\label{sec:physical}
Each agent occupies a context state $c\in\mathcal C$ and a corresponding
physical position $x=\Phi(c)$, where $\Phi:\mathcal C\to\mathbb R^2$
is a smooth projection determining how interaction modes are expressed 
in real space.  
The map $\Phi$ is not required to be injective: many distinct context 
states may share the same physical image, reflecting that $\mathcal C$ 
encodes only interaction-relevant degrees of freedom.

For a small context displacement $\delta c\in T_c\mathcal C$ (a tangent vector at $c$), the induced
physical displacement is
\begin{equation}
    \delta x = D\Phi(c)\,\delta c,
\end{equation}
where $D\Phi(c)$ is the Jacobian of the embedding.  
A nontrivial kernel of $D\Phi(c)$ implies that some context directions 
produce no physical motion, while others may be faithfully or 
anisotropically transmitted.  
Moreover, isolated points $c_*$ with $D\Phi(c_*)=0$ act as \emph{critical 
points} of the embedding: all first-order context displacements collapse 
to zero physical displacement, providing natural locations for vortex 
cores, pinning sites, or localized geometric defects in the induced flow.

A simple and instructive example is the projection
\begin{equation}
    \Phi(\theta_1,\theta_2) = (\cos\theta_1,\;\sin\theta_1),
\end{equation}
mapping the two-torus $T^2$ onto the unit circle in $\mathbb R^2$.  
This map is highly non-injective ($\theta_2$ is discarded), yet has a
rank--$1$ Jacobian almost everywhere:
\begin{equation}
    D\Phi(\theta_1,\theta_2)
    =
    \begin{pmatrix}
        -\sin\theta_1 & 0 \\
        \phantom{-}\cos\theta_1 & 0
    \end{pmatrix},
\end{equation}
so only motions in the $\theta_1$ direction generate physical motion, while
displacements in $\theta_2$ lie in the kernel of $D\Phi$.  
This illustrates how non-injectivity and partial rank deficiency arise
naturally when projecting $T^2$ into $\mathbb R^2$, and how geometric
filtering of context dynamics occurs through $D\Phi$.

The intrinsic context velocity generated by the Chern-Simons interaction, $v$, is pushed forward into physical space via
\begin{equation}
    v_{\mathrm{phys}}(c)= \dot x = D\Phi(c)\,v(c).
\end{equation}
Thus the geometry of $\Phi$ modulates and may selectively amplify,
attenuate, or localize the fundamentally chiral CS dynamics.  
Spatial variation in $D\Phi$ induces geometric nonreciprocity, while
critical points suppress motion and shape coherent structures such as 
vortices or density accumulation.

These analytic considerations motivate the numerical constructions 
developed in the next section, where specific embeddings $\Phi:T^2\to\mathbb R^2$
are chosen to probe chiral flows, nonreciprocity, and emergent collective
motion in physical space.

\subsection*{Operator Form of the Context-Physical Mapping}

For numerical implementation and for defining physical observables, it is
useful to express the pushforward of context fields in an operator form.
Rather than working directly with the pointwise Jacobian $D\Phi(c)$, we
introduce linear kernels
\[
M(x,c)\in\mathbb R^{2\times 2},
\qquad
W(x,c)\in\mathbb R,
\]
such that $M(x,\theta)= W(x,\theta)\, D\Phi(\theta).$
These map context--space velocity and density fields into their physical
counterparts.  These are defined by
\begin{align}
    v_{\mathrm{phys}}(x)
    =
    \int_{\mathcal C} M(x,c)\,v(c)\,d^2c,
    \\
    \rho_{\mathrm{phys}}(x)
    =
    \int_{\mathcal C} W(x,c)\,\rho(c)\,d^2c.
    \label{eq:MW-mapping}
\end{align}

The kernels $(M,W)$ provide a flexible representation of how context
dynamics are expressed physically.  The geometric pushforward associated
with $\Phi$ is recovered as the special case
\begin{align}
    M(x,c)&=\delta\!\bigl(x-\Phi(c)\bigr)\,D\Phi(c),
    \\
    W(x,c)&=\delta\!\bigl(x-\Phi(c)\bigr),
    \label{eq:MW-delta}
\end{align}
so that \eqref{eq:MW-mapping} reduces to
$v_{\mathrm{phys}}(x)=D\Phi(c)\,v(c)$ and
$\rho_{\mathrm{phys}}(x)=\rho(c)$ whenever $x=\Phi(c)$.
More general choices of $(M,W)$ allow for coarse–graining, 
non-injective projections, anisotropic filtering, or 
smoothing of fine-scale context features before they appear in physical
space.

In the spectral representation, \eqref{eq:MW-mapping} becomes
\begin{align}
    v_{\mathrm{phys}}(x)
    =
    \frac{1}{(2\pi)^2}
    \int_{\mathbb R^2}
    \widehat{M}(x,k)\,\widehat{v}(k)\,d^2k,
    \\
    \rho_{\mathrm{phys}}(x)
    =
    \frac{1}{(2\pi)^2}
    \int_{\mathbb R^2}
    \widehat{W}(x,k)\,\widehat{\rho}(k)\,d^2k,
\end{align}
where 
\begin{align}
\widehat{M}(x,k)
=
\int_{\mathcal C} M(x,c)\,e^{i k\cdot c}\,d^2c,
\\
\widehat{W}(x,k)
=
\int_{\mathcal C} W(x,c)\,e^{i k\cdot c}\,d^2c.
\end{align}
Thus, the geometry encoded in $\Phi$ may be implemented either through
its Jacobian $D\Phi(c)$ or through the associated operator kernels
$(M,W)$, which are particularly convenient for context dynamics evolved
in Fourier space.

\section{Numerical Simulation}
\label{sec:numerical}

Next, we consider the numerical simulation over a two-torus $T^2$ as the context manifold which maps to a two dimensional space $\mathbb R^2$. 
For computation, we choose a radius–modulated circular embedding defined by
\begin{align}
\Phi(\theta_1,\theta_2)
&=
\begin{pmatrix}
(R_0 + a \cos\theta_2)\cos\theta_1\\[4pt]
(R_0 + a \cos\theta_2)\sin\theta_1
\end{pmatrix},
\\
D\Phi(\theta)
&=
\begin{pmatrix}
\partial_{\theta_1}\Phi_x & \partial_{\theta_2}\Phi_x\\[4pt]
\partial_{\theta_1}\Phi_y & \partial_{\theta_2}\Phi_y
\end{pmatrix}.
\end{align}
We also choose 
\begin{align}
W(x,\theta)
=
\frac{
\exp\!\left( -\frac{1}{2\sigma^{2}}\, \| x - \Phi(\theta) \|^{2} \right)
}{
\displaystyle
\int_{\Omega}
\exp\!\left( -\frac{1}{2\sigma^{2}}\, \| x' - \Phi(\theta) \|^{2} \right)
\, dA(x')
},
\end{align}
where $dA$ means the area measure on the physical 2-dimensional domain.
The kernel satisfies the normalization property
\begin{align}
\int_{\Omega} W(x,\theta)\, dA(x) = 1
\quad\text{for each }\theta.
\end{align}

The distinct roles of $\Phi$ and $W$ clarify how contextual structure is translated into physical observables. The specific form of the embedding $\Phi(\theta_1, \theta_2)$ is chosen as the simplest construction that preserves angular structure while allowing internal degrees of freedom to modulate physical position. One contextual angle generates global rotation in physical space, whereas the second modulates the radius, embedding a two-dimensional toroidal context into $\mathbb{R}^2$ without enforcing a one-to-one correspondence. This choice introduces controlled non-injectivity, overlap, and position-dependent sensitivity purely through geometry, without invoking additional forces or interaction rules. The weighting kernel $W(x,\theta)$ then determines how this geometrically organized context is expressed, setting the spatial scale, smoothness, and degree of overlap through which contextual contributions are combined. Together, $\Phi$ fixes where and in what geometric manner context appears in physical space, while $W$ controls the resolution with which it is observed. Physical observables therefore arise from a weighted superposition of geometrically arranged context states, making collective structure, anisotropy, and apparent nonreciprocity natural consequences of the chosen embedding and weighting.

\begin{figure*}[!ht]
    \centering
    \includegraphics[width=0.45\linewidth]{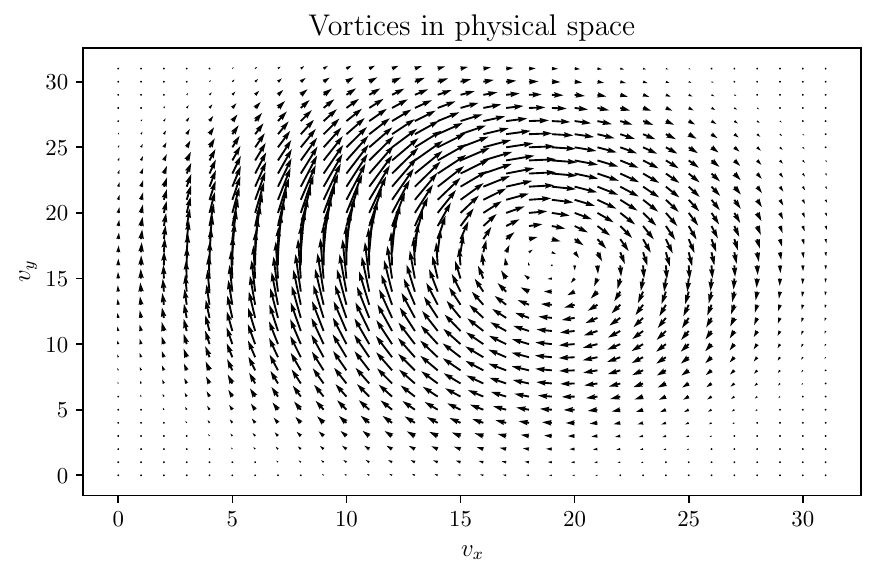}
    \includegraphics[width=0.45\linewidth]{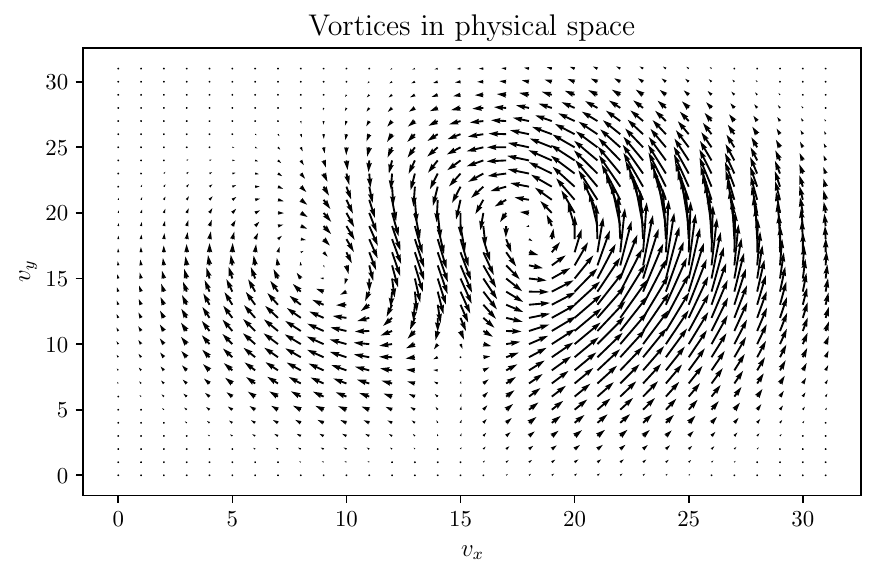}
    \caption{Final velocity distributions in physical space for two sample runs with $\gamma=3.0$.}
    \label{fig:vortices}
\end{figure*}

\subsection{Quantities Evaluated in Simulation}

The operator kernels $M(x,c)$ and $W(x,c)$ determine how context fields are
expressed in physical space:
\begin{align}
v_{\mathrm{phys}}(x,t)
&=
\int_{\mathcal C} M(x,c)\,v(c,t)\,dc,\\
\rho_{\mathrm{phys}}(x,t)
&=
\int_{\mathcal C} W(x,c)\,\rho(c,t)\,dc.
\end{align}
These maps allow non-injective projections, smoothing, or anisotropic
filtering of context dynamics before they appear in physical space. 
The degree of reciprocity is measured by
\begin{align}
\mathcal R(c,c')
&=
\frac{
\big\|
M(\cdot,c)\,\mathcal K(c,c')
+
M(\cdot,c')\,\mathcal K(c',c)
\big\|_F
}{
\|\mathcal K(c,c')\|_F + \varepsilon_0
}.
\label{eq:reci}
\end{align}
Values $\mathcal R>0$ indicate violation of reciprocity after mapping
into physical space, typically enhanced where $M$ is anisotropic or
rank-deficient.


The degree of coherent physical motion is quantified by
\begin{align}
P(t)
&=
\frac{
\big\|
\displaystyle\int_{ X}
\rho_{\mathrm{phys}}(x,t)\,v_{\mathrm{phys}}(x,t)\,dx
\big\|
}{
\displaystyle\int_{X}
\rho_{\mathrm{phys}}(x,t)\,\|v_{\mathrm{phys}}(x,t)\|\,dx
+\varepsilon_0
}.
\end{align}
Here $P(t)\simeq 1$ indicates strong collective alignment,  
while $P(t)\simeq 0$ indicates disordered or vortex-dominated flow.
Given the physical velocity field $v_{\mathrm{phys}}(x,t)$, the associated vorticity is defined by
\begin{align}
\omega(x,t)
&=
(\nabla_x \times v_{\mathrm{phys}})(x,t) \nonumber\\[4pt]
&=
\partial_x v_y(x,t) - \partial_y v_x(x,t).
\end{align}
We also define a useful quantity, enstrophy as the total rotational energy of the flow:
\begin{align}
E(t)
&=
\frac{1}{2}
\int_{\Omega} \omega(x,t)^2\, dA .
\end{align}
Enstrophy vanishes only for perfectly irrotational (gradient) flows and therefore provides a sensitive indicator of persistent rotational structure generated by non-reciprocal interactions.

\begin{figure*}[!ht]
    \centering
    \includegraphics[width=0.45\linewidth]{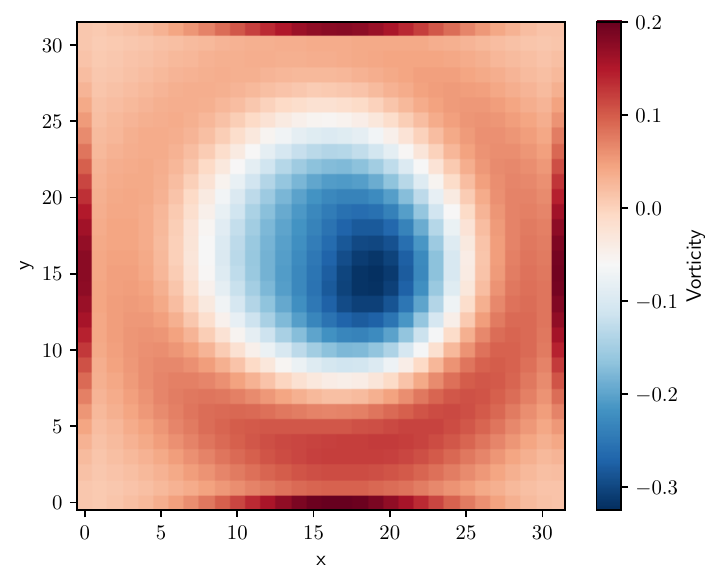}
    \includegraphics[width=0.45\linewidth]{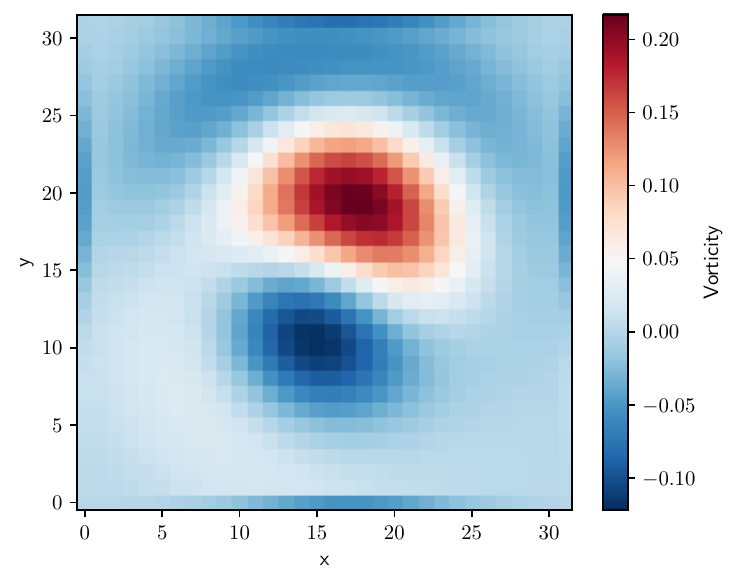}
    \caption{Final vorticity distributions in physical space for two sample runs with $\gamma=3.0$.}
    \label{fig:vorticity}
\end{figure*}

We quantify rotational activity in the physical flow using three circulation-based measures. The vorticity is $\omega(x,t) = (\nabla_x \times v_{\mathrm{phys}})(x,t)$, and the absolute total circulation,
\begin{align}
\Gamma_{\mathrm{abs}}(t) = \int_{\Omega} |\omega(x,t)|\, dA,
\end{align}
measures the overall magnitude of rotational motion independent of direction. Individual vortex circulations are computed by integrating vorticity over each detected vortex core $D_i$,
\begin{align}
\Gamma_i(t) = \int_{D_i} \omega(x,t)\, dA,
\end{align}
where the sign of $\Gamma_i$ indicates rotation direction and $|\Gamma_i|$ reflects vortex strength. The ensemble chirality is summarized by the Signed Circulation Dominance (SCD),
\begin{align}
\mathrm{SCD}(t) = 
\frac{\bigl|\sum_i \Gamma_i(t)\bigr|}{\sum_i |\Gamma_i(t)|},
\end{align}
which lies in $[0,1]$ and distinguishes balanced vortex populations ($\mathrm{SCD}\approx0$) from strongly chiral states where one rotational direction dominates ($\mathrm{SCD}\approx1$). Intuitively, $\Gamma_{\mathrm{abs}}$ captures how much rotation exists, while $\Gamma_i$ and SCD describe how that rotation is distributed and whether it exhibits a preferred handedness. We also study the hysteresis effect in mean alignment $P(t)$, $\mathrm{SCD}$, Enstrophy, and $\mathrm{Maximum Vorticity}$ with forward and backward sweeps of  $\gamma$.

\subsection{Results and Discussion}

Each simulation is performed on a $128\times128$ context grid ($N=128$) and mapped to a $64\times64$ physical domain ($N_x=64$) of size $L_p=2\pi$. The timestep is $\Delta t = 10^{-4}$, and trajectories are integrated for $2000$ steps with measurements recorded every $100$ steps. The diffusion coefficient is $D_c = 5\times10^{-5}$, and the Chern--Simons coupling parameters are $\gamma$ and $\kappa=1.0$, producing an effective curl strength $C = 2\pi\gamma/\kappa$. We vary $\gamma$ in the range $[0.5,6.0]$ in increments of $0.5$. For each value of $\gamma$, we generate an ensemble of $16$ independent simulations using different random initial conditions, and ensemble-averaged observables are reported. The embedding into physical space uses a radius $R_0=1.0$ with modulation amplitude $a=0.35$, and the context-to-physical projection employs a Gaussian kernel of width $\sigma = 0.15\times 2\pi$, normalized per context cell. A two-thirds spectral cutoff is applied to all Fourier-space operations to control aliasing. Together, these parameters define the geometry, non-reciprocal interaction strength, and numerical resolution used in the study.

\begin{figure*}[!ht]
    \centering
    \includegraphics[width=0.45\linewidth]{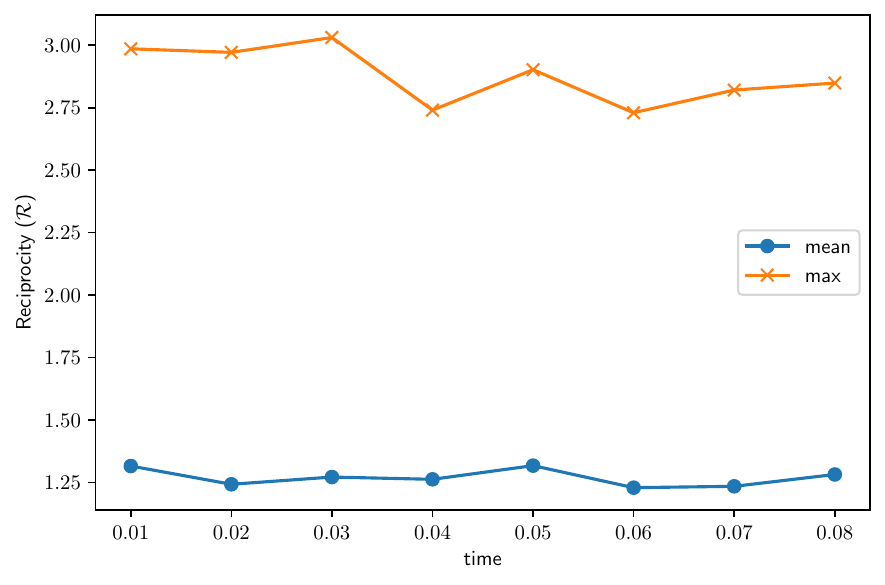}
    \includegraphics[width=0.45\linewidth]{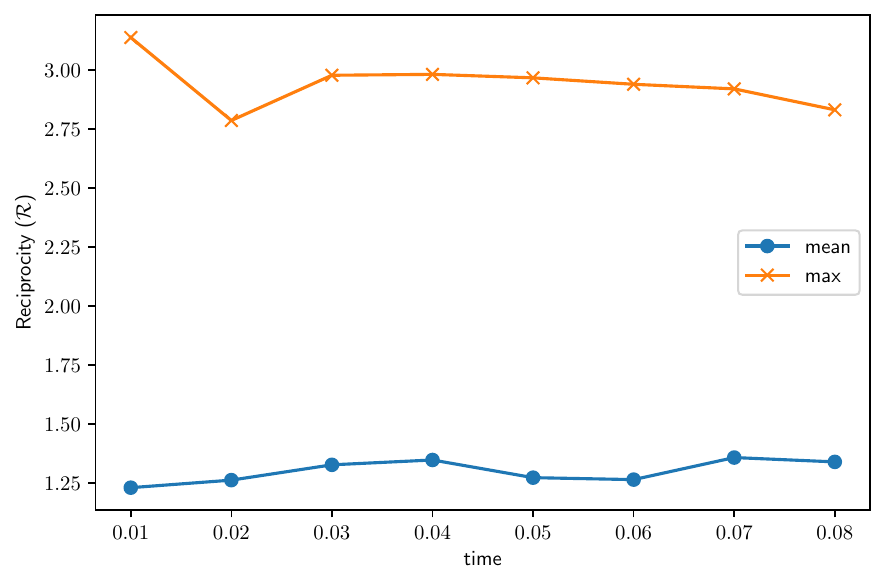}
    \caption{Final reciprocity distributions in physical space for two sample runs with $\gamma=3.0$. Reciprocity values greater than zero indicates non-reciprocal interaction.}
    \label{fig:reciprocity}
\end{figure*}

Figures~\ref{fig:vortices} and~\ref{fig:vorticity} show the final velocity and vorticity fields in physical space for two independent simulations at $\gamma = 3.0$. In both cases the flow organizes into coherent rotational structures, demonstrating the emergence of stable vortices driven by the antisymmetric Chern--Simons interaction. As predicted by the theory, the non-reciprocal coupling generates a velocity response proportional to a rotated gradient, producing a non-vanishing curl even in steady state. This manifests as localized circulation patterns in the velocity quiver plots (Fig.~\ref{fig:vortices}) and as pronounced regions of positive and negative vorticity (Fig.~\ref{fig:vorticity}). Although the detailed vortex positions differ across realizations due to stochastic initial conditions, the overall presence of persistent rotational motion is robust and characteristic of the intrinsic chiral dynamics induced by the Chern--Simons kernel.

Figure~\ref{fig:reciprocity} shows the sampled reciprocity residuals for two independent runs at $\gamma=3.0$, quantified using the operator-based measure $\mathcal{R}(c,c')$ defined in Eq.~\ref{eq:reci}. Both the mean and maximum residuals remain consistently above zero throughout the evolution, demonstrating persistent violation of reciprocity after mapping the Chern--Simons interaction into physical space. This behavior is a direct signature of the antisymmetric structure of the Chern--Simons kernel, which breaks action--reaction symmetry and generates intrinsically non-reciprocal flow responses. While fluctuations occur due to differences in the evolving density distribution, the overall magnitude and robustness of the residual indicate that non-reciprocity is not transient but an inherent property of the induced dynamics.

\begin{figure*}[!ht]
    \centering
    \includegraphics[width=0.45\linewidth]{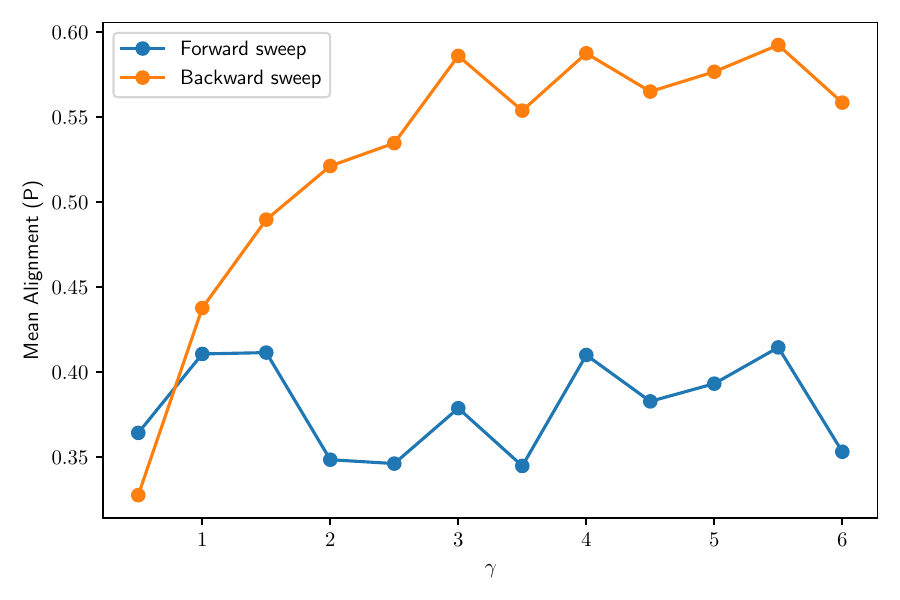}
    \includegraphics[width=0.45\linewidth]{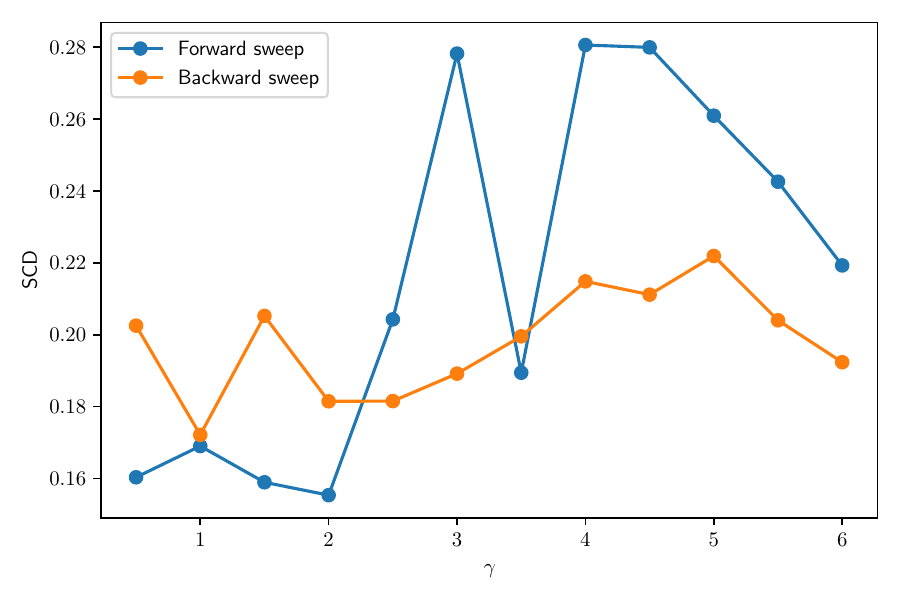}
    \includegraphics[width=0.45\linewidth]{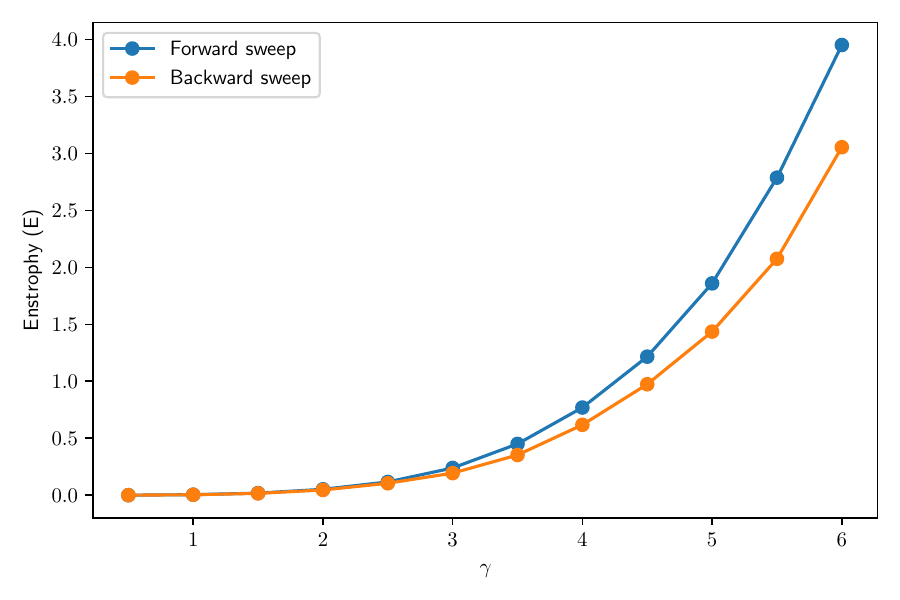}
    \includegraphics[width=0.45\linewidth]{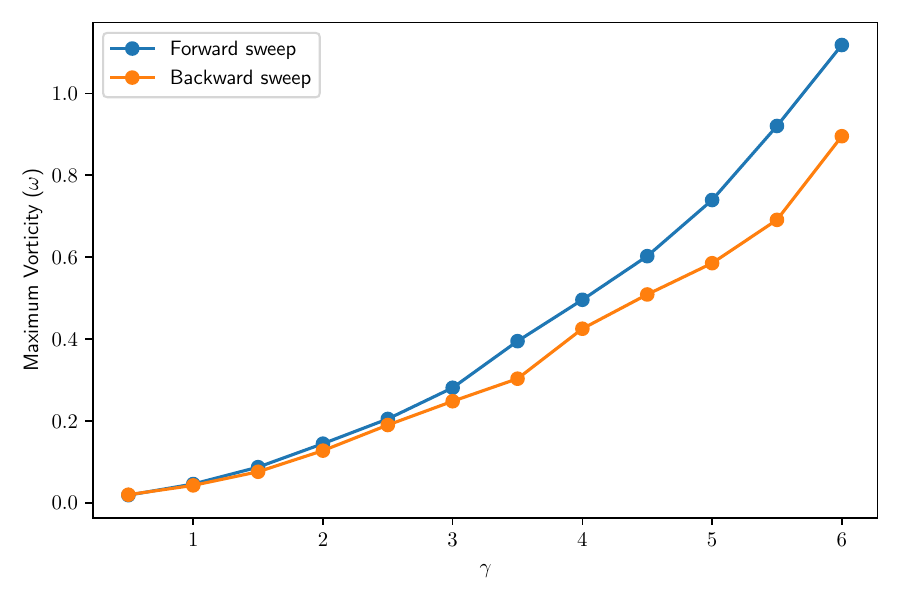}
    \caption{The variation of Mean Alignment $\mathrm{P}$, Signed Circulation Dominance (SCD), Enstrophy $\mathrm{E}$, and Maximum Vorticity $\omega$ with the forward and backward sweeps of $\gamma$. Forward and backward sweeps do not yield same trajectories and give rise hysteresis effect.}
    \label{fig:hysteresis}
\end{figure*}

Figure~\ref{fig:hysteresis} shows the hysteresis phenomenon obtained by sweeping the coupling parameter $\gamma$ forward and backward. Conceptually, $\gamma$ controls the strength of the feedback between internal geometry and motion. Because the gauge field depends self-consistently on the evolving context density, changing $\gamma$ does more than rescale velocities. It tunes how strongly motion reshapes the geometric field that drives it. This makes $\gamma$ a genuine control parameter: increasing it can lock the system into coherent, history-dependent collective states, while decreasing it weakens this feedback and leads to disordered dynamics. The observed hysteresis under forward and backward sweeps arises precisely because $\gamma$ regulates the activation of geometric back-action rather than acting as a simple force magnitude.

For all measured quantities such as mean alignment, signed circulation dominance (SCD), enstrophy, and maximum vorticity—the forward and backward branches clearly diverge, demonstrating strong path dependence in the emergent dynamics. Such hysteresis cannot occur in a reciprocal or potential-driven system, which would always return to the same state for a given value of $\gamma$. Instead, the observed loop structure reflects the intrinsically non-reciprocal Chern--Simons interactions. The physical-space response retains memory of how the system was driven, leading to distinct steady states depending on the sweep direction. The widening of the gap at larger $\gamma$ further indicates that stronger non-reciprocal coupling enhances this memory effect and stabilizes different chiral flow configurations on the forward and backward paths. It is worth mentioning that the mean reciprocity parameter $\mathcal{R}$ does not vary appreciably with $\gamma$. Thus, we have not included that in the figure.

\section{Conclusion}
\label{sec:conclusion}
We have shown that nonreciprocal collective dynamics can emerge from topology alone when active agents interact through a Chern--Simons gauge field defined on a compact internal context manifold.  The resulting Gauss-law constraint produces an intrinsically transverse, antisymmetric coupling, while the dynamical elimination of the gauge field generates higher-order nonlinear corrections that break action--reaction symmetry.  When mapped into physical space through an embedding or coarse-graining operator, these interactions give rise to robust chiral structures, persistent vorticity, finite circulation, and path-dependent transitions that manifest as clear hysteresis loops.  Our simulations demonstrate that these signatures persist across ensembles and parameter sweeps, indicating that the mechanism does not rely on fine tuning but is a generic consequence of the gauge topology.

The present framework identifies Chern--Simons flux as a minimal and universal mechanism for generating directional influence in active systems, offering a principled alternative to phenomenological non-Hermitian descriptions. More broadly, it establishes internal topology, not external fields or explicit asymmetries, as a fundamental driver of irreversible collective behaviour. This perspective invites new theoretical and experimental explorations of how geometric and topological structure in hidden internal coordinates can control macroscopic flow, pattern formation, and decision dynamics across biological, physical, and engineered active matter.

\bibliographystyle{unsrt}
\bibliography{ref}

\appendix
\section{Chern--Simons equation of motion and the origin of the current coupling}
\label{sec:Ampere-like-law}
The dynamical coupling between the gauge field and the agent current
follows directly from the variation of the full action.  The
Chern--Simons action in temporal gauge $a_0 = 0$ reduces to
\begin{equation}
    S_{\rm CS}
    =
    \frac{\kappa}{4\pi}
    \int 
    \epsilon^{ij}\,
    a_i \, \partial_t a_j
    \, d^2c\, dt ,
\end{equation}
which contains the only time derivative of the gauge potential.  Agents
contribute a conserved current $j^\mu = (\rho, j^i)$ and couple minimally
to the gauge field through
\begin{equation}
    S_{\rm coup}
    =
    -\int a_\mu j^\mu\, d^2c\, dt
    =
    -\int a_i\, j^i \, d^2c\, dt ,
\end{equation}
where we used $a_0 = 0$.  The total action is 
$S = S_{\rm CS} + S_{\rm coup}$.

To obtain the gauge-field equation of motion, we vary $S$ with respect to
$a_i$.  The variation of the Chern--Simons term is
\begin{align}
    \delta S_{\rm CS}
    &=
    \frac{\kappa}{4\pi}
    \int
    \epsilon^{ij}
    \left(
        \delta a_i\, \partial_t a_j
        +
        a_i\, \partial_t \delta a_j
    \right)
    d^2c\, dt \nonumber \\
    &=
    \frac{\kappa}{2\pi}
    \int 
    \epsilon^{ij}\,
    \delta a_i\, \partial_t a_j
    \, d^2c\, dt ,
\end{align}
where the second term has been integrated by parts in time.
The variation of the coupling term is simply
\begin{equation}
    \delta S_{\rm coup}
    =
    -\int \delta a_i \, j^i \, d^2c\, dt .
\end{equation}

Combining the two contributions gives
\begin{equation}
    \delta S
    =
    \int
    \delta a_i
    \left[
        \frac{\kappa}{2\pi}\,
        \epsilon^{ij}\partial_t a_j
        + j^i
    \right]
    d^2c\, dt .
\end{equation}
Requiring $\delta S = 0$ for arbitrary variations $\delta a_i$ yields the
Chern--Simons equation of motion,
\begin{equation}
    \frac{\kappa}{2\pi}\,
    \epsilon^{ij}\partial_t a_j
    + j^i = 0 ,
\end{equation}
or equivalently,
\begin{equation}
    -\,\frac{\kappa}{2\pi}\,
    \epsilon^{ij}\partial_t a_j
    = 
    j^i
    \;=\;
    \rho\, v^i .
    \label{CScurrent-final}
\end{equation}

Equation~\eqref{CScurrent-final} expresses how the gauge potential
responds dynamically to the spatial current.  This is similar to the Ampère-like law in electromagnetism. Unlike Maxwell electrodynamics, the Chern–Simons gauge field has no propagating wave modes: the equations of motion are first order, and the gauge potential is slaved instantaneously to the agent density and current. As a result, context fluctuations do not propagate independently; their evolution is entirely determined by the motion of the agents themselves.

Together with the overdamped
relation $v^i = -\gamma a^i$, it provides the feedback mechanism
responsible for the emergence of non-reciprocal interactions in the full
Chern--Simons dynamics.

\end{document}